# Plagiarism: Taxonomy, Tools and Detection Techniques

Hussain A Chowdhury and Dhruba K Bhattacharyya

Dept. of CSE, Tezpur University

**Abstract**

To detect plagiarism of any form, it is essential to have broad knowledge of its possible forms and classes, and existence of various tools and systems for its detection. Based on impact or severity of damages, plagiarism may occur in an article or in any production in a number of ways. This survey presents a taxonomy of various plagiarism forms and include discussion on each of these forms. Over the years, a good number tools and techniques have been introduced to detect plagiarism. This paper highlights few promising methods for plagiarism detection based on machine learning techniques. We analyse the pros and cons of these methods and finally we highlight a list of issues and research challenges related to this evolving research problem.

1. **Introduction**

   Due to the digital era, the volume of digital resources has been increasing in the World Wide Web tremendously. Today, creation of such digital resources and their storage and dissemination are simple and straight forward. With the rapid growth of these digital resources, the possibility of copyright violation and plagiarism has also been increasing simultaneously. To address this issue, researchers started working on plagiarism detection in different languages since 1990. It was pioneered by a copy detection method in digital documents [1]. However, the software misuse detection was initiated even much earlier, in 1970 by detecting plagiarism among programs [2]. Since then, a good number of methods and tools have been developed on plagiarism detection which are available online. But it is very much chaotic when one wants to choose the best plagiarism detection method or plagiarism detection tool. It may be due to lack of controlled evaluation environment in plagiarism detection research. Plagiarism is the presentation of another's words, work or idea as one's own [3]. It has two components, viz., (1) Taking the words, work or ideas from some source(s) and (2) Presenting it without acknowledgments of the source(s) from where words, works or ideas are taken [3]. Plagiarism can appear in different forms in an articles. However, there are mainly two types of plagiarisms typically found to occur, such as (1) textual plagiarism and (2) source code plagiarism [4]. Plagiarism may occur within same natural language or it may appear between two or more different languages. Many researchers or software companies still trying to provide an efficient method or tool for plagiarism detection. There are mainly two types of plagiarism detection approaches available based on whether external resources or references are used or not during plagiarism detection, such as (1) intrinsic plagiarism detection, where no external references are used and (2) extrinsic plagiarism detection, where external references are used [5].

In the yester years, a good number of tools and techniques have been introduced to detect plagiarism of various forms. Several efforts have been made [6] [5][7][8] to survey these works. However, unlike other surveys, this survey is attractive in view of the following points.

- It reports a comprehensive and systematic survey on a large number of methods of plagiarism detection and analyzes their pros and cons.
- It includes discussion on a large number of tools on plagiarism detection and reports their features. It also compares these tools based on a set of crucial parameters.
- Finally, in includes a list of issues and research challenges.

The rest of the paper is organized as follows: Section 2 presents fundamentals of plagiarism and its classification. It reports a taxonomy of various known plagiarism forms. In section 3 we discuss fundamentals of detection and a large number of detection techniques. Section 4 reports a list of tools for plagiarism detection and discuss their features. Section 5 presents a list of issues and research challenges. Finally, section 6 draws the conclusions of this paper.

## 2. Plagiarism and Its Types

As stated in [3], plagiarism can be defined as an appropriation of the ideas, words, process or results of another person without proper acknowledgment, credit or citation. Plagiarism can appear in a research article or program in following ways:
a. Claiming another person's work as your own.
b. Use of another person's work without giving credit.
c. Majority of someone's contribution as your own, whether credit is given or not.
d. Restructuring the other works and claiming as your own work.
e. Providing wrong acknowledgment of other works in your work.

Plagiarism can appear in different forms in a document, work, production or program. Two basic types of plagiarisms [9] are (a) Textual plagiarism and (b) Source Code plagiarism.

> The nearest-neighbour based outlier mining technique is able to detect a plagiarized text segment.(*Active voice*)
>
> A plagiarized text segment is able to detect by the nearest-neighbour based outlier mining technique.(*passive voice*)

(a)

```
Data: First, Last              Data: Start, Finish
Result: Sum                    Result: Total
while (Last ≠ 0) do            while (Finish ≠ 0) do
    Sum=First*Last;                Total=Start*Finish;
    Last=Last-1;                   Finish=Finish-1;
end                            end
```
(b)

Figure 1: Examples of (a) Textual (b) Source Code Plagiarism

*Textual plagiarism* is commonly seen in education and research. Figure 1 (a) shows an example of textual plagiarism where entire word-for-word are taken from source without direct quotation. Textual plagiarism further can be divided into seven sub categories based on its forms and application [4][10] as shown in Figure 2. We discuss each of these in brief, next.
1. *Deliberate copy-paste/clone plagiarism:* This type of textual plagiarism refers to copying other works and presenting as if your own work with or without acknowledging the original source.
2. Paraphrasing plagiarism: This form of plagiarism can occur on two ways as given below.
    - Simple paraphrasing: It refers to use of other idea, words or work, and presenting it in different ways by switching words, changing sentence construction and changing in grammar style.
    - Mosaic/Hybrid/patchwork paraphrasing: This form of textual plagiarism generally occurs when you combine multiple research contributions of some others and present it in a different way by changing structure and pattern of sentence, replacing words with synonyms and by applying a different grammar style without citing the source(s).
3. *Metaphor plagiarism:* Metaphors are used to present others idea in a clear and better manner.
4. *Idea plagiarism:* Here, idea or solution is borrowed from other source(s) and claiming as your own in a research paper.
5. *Self/recycled plagiarism:* In this form, an author uses his/her own previous published work in a new research paper for publication.
6. *404 Error / Illegitimate Source plagiarism:* Here, an author cites some references but the sources are invalid.
7. *Retweet plagiarism:* In this form an author cites the reference of proper source but his/her presentation is very similar in the scene of original content wordings, sentence structures and/or grammar usage.

Based on characteristics, plagiarism can also be categorized into *literal* and *intelligent* plagiarism. Literal plagiarism consists of copy-paste/clone, paraphrasing, self/recycled, and retweet plagiarism. The other form of plagiarism can be considered as intelligent type of plagiarism. In general, intrinsic plagiarism detection methods can detect paraphrasing, idea, and mosaic textual plagiarism sub-types [4], whereas, external plagiarism detection methods can detect clone, metaphor, retweet and possibly (with a low probability or none) self plagiarisms and error 404 [4]. However, in the recent developments, such demonstration is not very prominent between these two types of detection methods.

In *source code plagiarism*, codes written by others are copied or reused or modified or converted a part of codes and claimed as one's own. Figure 1 (b) shows the example an example of source code plagiarism where entire program is represented again in different way by changing syntax. Typically it is seen, in educational institutes. This type of plagiarism, as shown in Figure 2 can be divided into four subtypes.

1. *Manipulation from Vicinity plagiarism*: Here, a developer manipulates a program by (i) inserting, (ii) deleting, or (iii) substituting some codes in an existing program, with or without acknowledging the original source and claiming it as his/her own program.

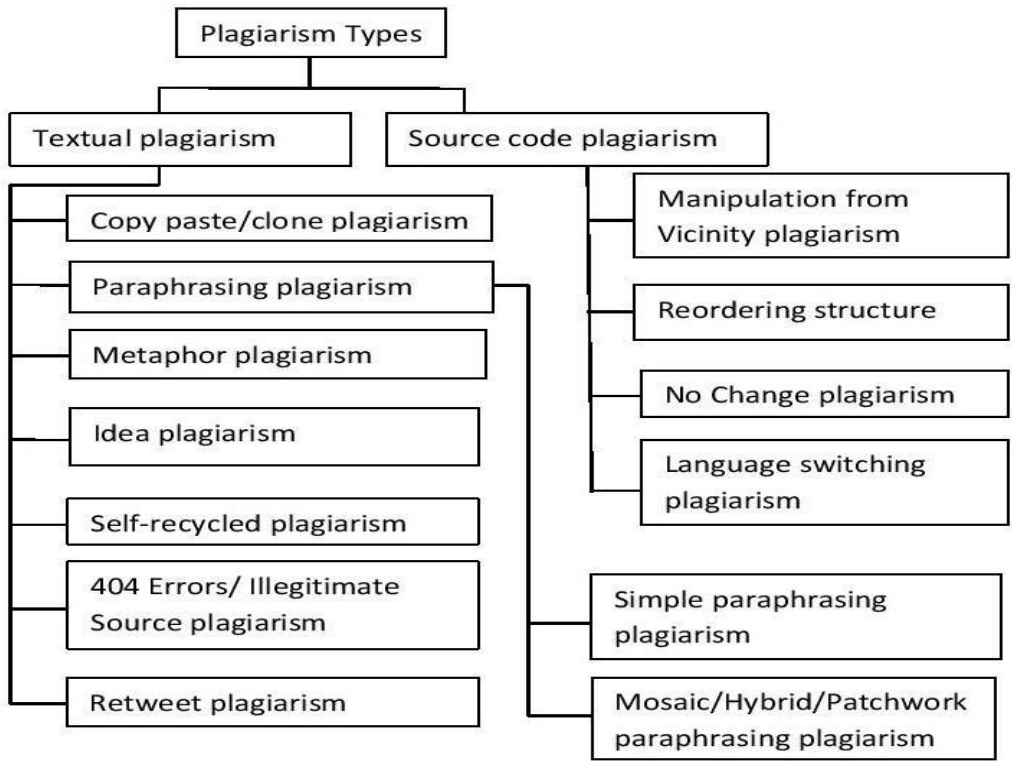

Figure 2: Taxonomy of plagiarism

2. Reordering structure plagiarism: In this type, the developer reorders the statements or functions of a program or changes syntax of a program without referring the original source.
3. No change plagiarism: Here, the developer adds or removes white spaces or comments or indentation of the program and claims the program as his/her own program.
4. Language switching plagiarism: In this type, the developer changes the languages, or a program written in one language is rewritten in another language and declares it as his/her own.

**3. Plagiarism Detection**

Plagiarism can occur between two same or two different natural languages. Based on language homogeneity or heterogeneity of the textual documents being compared, the plagiarism detection can be divided into two basic types [5] i.e., monolingual and cross-lingual.
1. *Monolingual Plagiarism Detection*: This type of detection deals with homogeneous language settings e.g., English-English. Most detection methods are of this category. It can be further divided into two subtypes based on the use of external references during detection.
    a) Intrinsic Plagiarism Detection: This detection approach analyses the writing style or uniqueness of the author and attempts to detect plagiarism based on own-conformity or deviation between the text segments. It does not require any external sources for detection.
    b) Extrinsic Plagiarism Detection: Unlike the intrinsic approach, this approach compares a submitted research article against many other available relevant digital resources in repositories or in the Web for detection of plagiarism. }

2. *Cross-Lingual Plagiarism Detection*: This detection approach is able to perform in heterogeneous language settings e.g., English-Chinese. There are only a few cross-lingual plagiarism detection methods available due to difficulty in finding proximity between two text segments for different languages.

In Figure 3 a schematic view of the basic plagiarism detection has been shown for both text documents and source codes. It accepts an input candidate text document and attempts to identify the text segments in the document plagiarized from some sources. It can be in monolingual as well as in cross-lingual framework. Textual plagiarism detection can be classified based on how textual features are used to characterize documents. There are different textual features like lexical feature, syntactic feature, semantic feature and structural feature, which can be used to detect similarity between two documents. These textual features are used in both extrinsic and intrinsic as well as in cross-lingual plagiarism detection.

In case of source code plagiarism detection, human intervention is required to detect plagiarism. Source code similarity detection can be carried out in various ways, such as (i) string matching, (ii) token matching, (iii) parse tree matching, (iv) program dependency graph (PDG) matching, (v) similarity-score matching and (vi) by hybridization of the above [11].

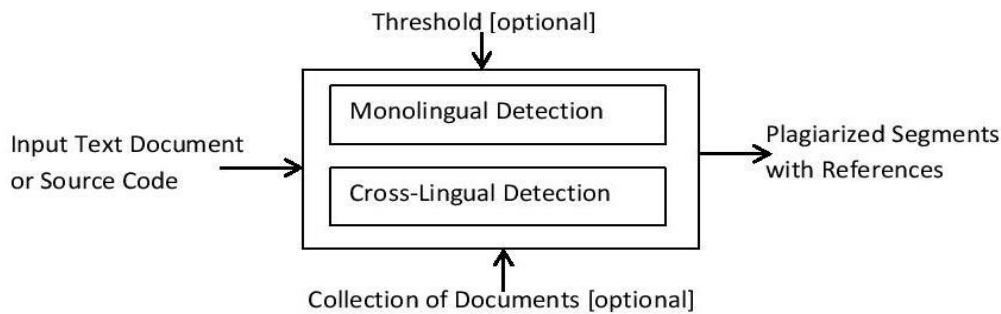

Figure 3: Basic plagiarism detection system

3.1 Similarity Measures for Comparing Documents or text segments

To detect plagiarism we have to measure similarity between two documents. We observe that most researchers use the following two types of similarity metrics.
1. *String Similarity Metric:* This method is commonly used by extrinsic plagiarism detection algorithms. Hamming distance is a well-known example of this metric which estimates number of characters different between two strings x and y of equal length, Levenshtein Distance [12][13] is another example, that defines minimum edit distance which transform x into y, similarly, Longest Common Subsequence [14][2] measures the length of the longest pairing of characters between a pair of strings, x and y with respect to the order of the characters.
2. *Vector Similarity Metric:* Over the decade, a good number of vector similarity metrics have been introduced. A vector based similarity metric is useful in calculating similarity between two different documents. Matching Coefficient [15] is such a metric that calculates similarity between two equal length vectors. Jaccard Coefficient [16] is author such metric used to define number of shared terms against total number of terms between two identical vectors, Dice Coefficient [17] is similar to Jaccard but it reduces the number of shared terms, Overlap Coefficient [18] can compute similarity in terms of subset matching, Cosine Coefficient [19] to find the cosine angle between two vectors, Euclidean Distance the geometric distance between two vectors, Squared Euclidean Distance places greater weight on that are further apart, and Manhattan Distance can estimate the average difference across dimensions and yields results similar to the simple euclidean distance.

## 3.2 Plagiarism Detection Methods

Detection of plagiarism in text document with high accuracy is a challenging task. In the past two decades, a large number of methods have been reported by researchers to handle this task. These methods can be classified into eleven distinct categories. Some prominent methods under each of these categories are discussed next. Also, we have analysed their pros and cons, and reported in a tabular form in Table 1.

1. *Character-Based Methods:* Most plagiarism detection methods belong to this category. These methods exploit character-based, word-based, and syntax-based features. It utilizes these features to find similarity between a query document and existing documents. However, the similarity between a pair of documents may be estimated using both exact matching and approximate matching. In exact matching, every letter in both the strings must be matched in the same order. Our survey reveals that most detection techniques are developed based on n-gram or word n-gram based exact string similarity finding approach. For instance, Grozea et al. [20] use character 16-gram matching, whereas the authors of [21] use word 8-gram matching. Similarly, some researcher has made an effective use of approximate string matching approach. This string matching shows degree of similarity/dissimilarity between two strings. There are several proximity measures available to support the approximate string matching. One can use string similarity metric or vector similarity metric for the purpose.

2. *Vector-Based Method:* Here, lexical and syntax features are extracted and categorized as tokens rather than strings. The similarity can be computed using various vector similarity measures like Jaccard, Dice's, Overlap, Cosine, Euclidean and Manhattan coefficients. Our observation is Cosine coefficient and Jaccard coefficients are popular and effective in finding similarity between two vectors. Cosine coefficient in detecting partial plagiarism without sharing documents content. Hence it is useful to detect plagiarism in documents where submission is considered as confidential [22].

3. *Syntax-Based Methods*: These methods exploit syntactical features like part of speech (POS) of phrase and words in different statements to detect plagiarism. The elements of basic POS tag are verbs, nouns, pronouns, adjectives, adverbs, prepositions, conjunctions and interjections. In [14] [13], the authors use POS tag features followed by string similarity metric to analyse and calculate similarity between texts. The authors of [23] use syntactical POS tag to represent a text structure as a basis for further comparison and analysis i.e., documents containing same POS tag features are carried out for further analysis and for identification of source of a plagiarism.

4. *Semantic-Based Methods:* A sentence may be defined as an ordered group of words. Two sentences may be same but the order of their words may be different. In Figure 1 (a), sentence is constructed by just transforming from active voice to passive voice but the semantics of the sentences are same. WordNet [24] is used in this content to find the semantic similarity between words or sentences. The degree of similarity between two words used in knowledge-based measures by Gelbukh [25] is calculated using information from a dictionary. This similarity between two words is used as semantic similarity between two words. In another approach, Resnik [26] used WordNet to calculate the semantic similarity, whereas, Leacock's et al., [27] determine semantic similarity by counting the number of nodes of shortest path between two concepts.

5. *Fuzzy-Based Methods:* In a fuzzy-based method, similarity of text such as sentences is represented by values ranging from zero (entirely different) to one (exactly matched). Here, the words in a documents are represented using a set of words of similar meaning and sets are considered as fuzzy since each word of the documents is associated with a degree of similarity [28]. This method is attractive because it can detect similarity between documents with uncertainty. In [28], a correlation matrix is constructed which consists of words and their corresponding correlation factors that measures the degree of similarity among different words. Then, it obtains the degree of similarity among sentences by computing the correlation factors between pair of words from two different sentences in their respective documents. In [29], the degree of similarity of two documents or any two Web documents are identified by using fuzzy IR approach. The authors introduce a tool for this purpose. There is another method discussed in [30] which adapts fuzzy approach to find in what extent two Arabic statements are similar. For that they used a plagiarism corpus of 4477 sources statements and 303 query/suspicious statements.
6. *Structure-Based Methods:* Unlike those methods above, developed based on lexical, syntactic, and semantic features of the text in documents to find similarity between two documents, a structure based method uses contextual similarity such as how the words are used in entire documents. However, our survey can find a few methods of this category. Contextual information is generally handled using tree-structure feature representation as can be found in ML-SOM [31]. In [32], the author detects plagiarism in two steps. First step performs document clustering and candidate retrieval using tree-structure feature representation and second step detects by utilizing ML-SOM.
7. *Stylometric-Based Methods:* These methods aim to quantify the writing styles of the author to detect plagiarism. It computes, similarity score between two sections or paragraphs or sentences based on stylometric features of the authors. These methods are instances of intrinsic plagiarism. The style representation formula may be writer specific or reader specific [33]. A writer specific style is mostly with author's vocabulary strength or complexity of presenting a document. On the other hand, a reader specific style deals with how a reader can easily understand the texts. One can find usefulness of outlier mining to detect plagiarism in a document under this approach. A detail discussion on Stylometric-Based methods is available in [34].
8. *Methods for Cross-Lingual Plagiarism Detection*: Cross-lingual plagiarism detection is a challenging task. It requires in depth knowledge of multiple languages. Finding appropriate similarity metric for such method is also an important issue. This type of methods work based on cross-lingual text features. Various types of these methods include (1) cross-lingual syntax based methods, (2) cross-lingual dictionary based method, and (3) cross-lingual dictionary based methods [5]. A detail survey on Cross-Lingual methods is done in [35]. In [20], a statistical model is used to evaluate the similarity between two documents regardless of the order in which the terms appear in suspected and original documents [36].
9. *Grammar Semantics Hybrid Plagiarism Detection Methods:* These methods are effective method in plagiarism detection for their use of natural language processing. They are capable of detecting copy/paste and paraphrasing plagiarism accurately. Such methods eliminate the limitations of semantic-based method. A semantic-based method usually cannot detect and determine the location of plagiarised part of the document but such grammar-based method can address this issue efficiently [37][5].
10. *Classification and Cluster-Based Methods*: In information retrieval process, supervised and unsupervised grouping of documents plays an important role. In many research problem such as text summarization [38], text classification [39],

Table 1: PLAGIARISM DETECTION TECHNIQUES: A General Comparison

| Author & Name | Intrinsic(I)/Extrinsic | Approach used | model used | | Language(s) | Types of plagiarism | | | | | | | References |
|---|---|---|---|---|---|---|---|---|---|---|---|---|---|
| | | | Mono-Lingual IR | Cross-Lingual IR | | Literal | | Intelligent | | | | | |
| | | | | | | Copy | Near copy | Restructuring | Paraphrasing | Summarising | Translating | Idea(Section) | Idea(Context) |
| Character-Based (CNG) | E | String Matching | ✓ | | any | ✓ | ✓ | | | | | | |
| Vector-Based(VEC) | E | Text Similarity | ✓ | | any | ✓ | ✓ | ✓ | | | | | |
| Syntax-Based(SYN) | E | Text Similarity | ✓ | | specific | ✓ | ✓ | ✓ | | | | | |
| Semantic-Based(SEM) | E | Word Similarity and Local Semantic Density | ✓ | | specific | ✓ | ✓ | ✓ | ✓ | ? | | | |
| Fuzzy-Based(FUZZY) | E | Fuzzy set of synnomy words | ✓ | | specific | ✓ | ✓ | ✓ | ✓ | ? | | | |
| Structural-Based(STRUC) | E | Tree-Structured Features Representation | ✓ | | specific | ✓ | ✓ | ✓ | ? | ? | | ? | ? |
| Stylometric-Bsed(STYLE) | I | Author vocabulary richness and style complexity | ✓ | | specific | ✓ | ✓ | ✓ | | | | | |
| Cross-Lingual(CROSS) | E | Cross-Lingual Syntax, semantic, dictionary, statistic | | ✓ | cross | | ✓ | | | | ✓ | | |
| Grammar-Based(GRAM) | E | String Matching | ✓ | | any | ✓ | ✓ | ✓ | | | | | |
| Cluster-Based(CLUS) | E | Text summerization and exact matching | | | specific | ✓ | ✓ | ✓ | ✓ | | | | |
| Citation-Based(CITE) | E | Word Similarity and Local Semantic Density | ✓ | | specific | ✓ | ✓ | ✓ | ✓ | ? | | | |

Note: IR: Information Retrieval, ?: Need further research

and plagiarism detection [40], classification and clustering are useful in reducing the search space during the information retrieval process. It helps in reducing the document comparison time significantly during plagiarism detection. Some methods [41][42] use keywords or specific words to cluster the similar sections of documents.

11. *Citation-Based Methods:* In [43], a novel method is proposed to detect plagiarism in citation basis. This method is a new approach towards detecting plagiarism and scientific documents that have been read but not cited. Citation-based methods belong to semantic plagiarism detection techniques because these techniques use semantics contained in the citation in a document [44]. The similarity between two documents is computed based on the similar patterns in the citation sequences [44].

## 4. Plagiarism Detection Tools

In the past two decade, several plagiarism detection tools have been developed. Some of these tools are discussed in brief, next. Also, we have analyzed their pros and cons, and reported in a tabular form in Table 2 We reported the classification of tools in Figure 4

i. **SafeAssignment [6]:** This anti-plagiarism checker claims to search an index of 8 billion documents available in the Web. It uses some major scholastic databases like ProQuest™, FindArticles™ and Paper Mills during searching and detection process. SafeAssignment maintains a database where user account is essential to keep fingerprints of the submitted documents in order to avoid any legal or copy right problem. This tool uses proprietary searching and ranking algorithms for match detection of fingerprints with its resources. The results of plagiarism detection is presented to the user within couple of minutes.

ii. **Docol©c[6]:** This Web based service uses capabilities like searching and ranking of Google API. The submitted document is uploaded to a server and evaluation is done in the server side. The software provides a simple console to set fingerprint (search fragments) size, date constraints, filtering and other report related options. The evaluation result is sent to the user through email identifying plagiarized sections and sources of plagiarism. This is totally Google API dependent and so it may be unavailable at any point of time.

iii. **Urkund [6]:** This is another Web based service which carry out plagiarism detection in server side. This is an integrated and automated solution for plagiarism detection. This is a paid service which uses standard email system for document submission and for viewing results. This system claims to process 300 different types of document submissions and it searches through all available online sources. It gives more priority to educational sources of documents more during searching.

iv. **Copycatch [6]:** This is a client-based tool which utilizes the local database of documents during comparison. It offers 'gold' and 'campus versions', providing comparison capabilities against large repository of local resources. It has another Web version which utilizes the capabilities of Google API for plagiarism detection across the Internet. To use the Web version, user needs personal Google API licence through signup.

v. **WCopyfind[6]:** It is an open source plagiarism detection tool for detection of words or phrases of defined length within a local repository of documents. Its extended version has the capabilities of searching across the Internet using Google API to check plagiarism online.

vi. **Eve2 (Essay Verification Engine [6][45] :** This system is installed in user's computer and it checks plagiarism of a document against Internet sources. It does not contact any online database. It accepts text in several formats but internally converts the input file into text for processing. It presents the user with a report identifying matches found in the Web.

vii. **GPSP - Glatt Plagiarism Screening Program [6]:** This system uses different approaches unlike other mentioned services. It finds and uses the writing style of the author(s) to detect plagiarism. This service works locally and it asks the author to go through a test by filling the blank spaces. The number of correctly filled spaces and time taken to complete the test are used to make a hypothesis about plagiarism. This system is basically developed for teachers and it cannot detect source code plagiarism.

viii. **MOSS - a Measure of Software Similarity [46]:** This system is used to detect source code plagiarism. This service takes batches of documents as input and attempts to present a set of HTML pages to specify the sections of a pair of documents where matches detected. The tool specializes in detecting plagiarism in C, C++, Java, Pascal, Ada, ML, Lisp, or Scheme programs.

ix. **JPlag [47][6]:** It is a Web based source code plagiarism detection tool started in 1997. The tool accepts a set of programs as input to be compared and to present a report identifying matches. JPlag carry out programming language syntax and structure aware analysis to find results. It can detect plagiarism in Java, C and C++ programs. The execution time of this service is less than one minute for submissions of 100 programs of several hundred lines each.

x. **Copyscape [48]:** This system takes URL as input and search for copies of a Web page in the Internet. Copyscape helps to find sites that have copied from someone's Web page content without permission. It has both free and premium version and it pushes the free users to buy their premium by limiting the search features.

xi. **DOC Cop [49]:** This plagiarism detection system creates report displaying the correlation and matches between documents or between documents and the Web. It is free plagiarism detection system. \

xii. **Ephorus [46]:** To access this tool, user is to register with the Ephorus site. Hence, no downloads or installation is needed. The search engine compares a text document to millions of others on the Web and reports back with an originality report [50]. This tool can be freely tried but license needs to be purchased. It is well known in many European universities and organization.

xiii. **ithenticate [51][46]:** This is a successful Web based plagiarism detection tool for any text document. This tool is not required to install in client computer. This application compares input documents against the document sources available on the Web. This well-known tool is used by most well-known journal publishers. It is a easy to use, quick plagiarism checker for professionals. It is designed to be used by institutions rather than personal, but lastly they provided a limit service for single plagiarism detection user like master and doctoral students and this allows them to check a single document of up to 25,000 words.

xiv. **Plagiarism Detect [46]:** To use this tool, user needs to register by providing correct information. After registration, users are allowed to input text in a given text box or as a file by uploading for analysis. This is a free service which finally sends evaluation report to the user's email account with a list of links from where information are copied. It also specifies amount of plagiarism (in \%) detected. User needs to download and install the software in order to use it.

xv. **Exactus Like [52]:** This plagiarism detection system is not able to find simple copy-paste plagiarism but also can detect moderately disguised borrowing (word/phrase reordering, substitution of some words with synonyms [52]. To do this, the system leverages deep parsing techniques. This Web based tool supports most of the popular file formats such as Adobe PDF, Microsoft Word, RTF, ODT and HTML. Currently Exactus Like includes about 8.5 million indexed documents. Internally this tool is basically a distributed system and a demo version of this tool is available online.

xvi. **DupliChecker [53]:** It is a free online plagiarism checker. This tool can be accessed by unregistered user only once, but registered user can check for plagiarism for 50 times in a day. The input file must contain more than 1000 words per similarity

xvii. **Plagiarisma [54]:** search. User can check content's originality by number of ways such as via copy paste, uploading file or by submitting URL.

xvii. **Plagiarisma [54]:** It is free and simple plagiarism checking tool. This software supports 190+ languages and it does not store any scanned content. The input file can be provided in three ways (1) Copy paste (2) Check by entering URL and (3) Uploading file. However, the tool lacks of advanced features so it cannot be relied for heavy scanning works.

xviii. **Plagiarism Checker [48]:** It was first available in early 2006. This freely available online service uses Google or Yahoo service to check whether documents submitted by students are copied from Internet material or not. It simply encloses each phrase in quotation marks and inserts an OR between each phrase during checking.

xix. **Plagium [55][8]:** This simple plagiarism detection tool, is effective in comparison to many of its counterpart, both in terms of results and algorithm. Though Plagium can be used free to some extent using quick search, their paid version has added benefits such as timeline feature and alert feature which pops up whenever someone's content is plagiarised. This tool has flexibility in pricing option like we can buy search credit either as prepaid plans or monthly plans. This tool allows user to check for plagiarism upto 5000 words without signing up.

xx. **PlagTracker [56]:** It is a popular plagiarism checker for students, teachers, publishers and Website owners. It has a large database of academic publications in million and provides detail report of the scanned work. If someone wants to check assignments in bulk, it requires to subscribe monthly. This tool found useful to ensure whether a test document is plagiarized or not.

xxi. **Quetext [57]:** It uses Natural Language Processing and Machine Learning to detect plagiarism. It performs first internal plagiarism checking and then it goes for external checking. This free tool uses every possible factor for each word to detect plagiarism. It provides support to multiple languages and one can search for unlimited words. To check plagiarism with this tool, one needs just plain copy paste of the text document. The main disadvantage of this tool is that it does not provide detail report. Also it is not user friendly.

xxii. **Turnitin [46][58]:** This an another successful Web based tool provided by iParadigms. The user is needed to upload test document to the system database for plagiarism check the system creates a fingerprint of the document and stores it. In this tool, detection and report generation is carried out remotely. Turnitin is already accepted by 15,000 Institutions and 30 Million Students due to easy to use interface, support of large repository, detailed text plagiarism check and well organized report generation. It can be considered as one of the best plagiarism checkers for teachers.

xxiii. **Viper [8]:** This free plagiarism scanner scans the submitted documents against 10 billion sources and documents present in a computer. It gives peace of mind regarding any accidental plagiarism. This tool offers unlimited resubmitting of documents and it provides links to plagiarised work in the reports.

xxiv. **Maulik [59]:** Maulik detects plagiarism in Hindi documents. It divides the text into n-grams and then matches with the text present in the repository as well as with documents present online. It uses Cosine similarity for finding the similarity score. Maulik is capable of finding plagiarism if root of a word is used or a word is replaced by its synonyms. This tool is superior than existing. Hindi plagiarism detection tools such as Plagiarism checker, Plagiarism finder, Plagiarisma, Dupli checker, and Quetext.

xxv. **Plagiarism Scanner [8]:** This is a fast and effective plagiarism detection tool for students, instructors, publishers, bloggers since 2008. It is a user-friendly online tool. This tool conducts through an in-detail detection for plagiarism of a submitted document within a few minute only. This tool runs against all Internet resources, including Websites, digital databases, and online libraries (such as Questia, ProQuest, etc). It generates a full report, indicating the overall originality rating and the percentage of plagiarized materials in the submitted text. It also provides

Table 2: PLAGIARISM DETECTION TOOLS: A General Comparison

| Tool Name & Author | Year | Extrinsic(E)/Intrinsic(I) | User friendly | Submission of single/multiple Files? | Source code availability? | Source(Ref) |
|---|---|---|---|---|---|---|
| SafeAssignment by Mydropbox | 2008 | E | yes | single | No(Free) | http://www.safeassign.com/ |
| Docolc by IFALT9 | 2005 | E | yes | single | No(Free) | https://www.docoloc.de/ |
| Urkund by group of teachers | 2000 | E | yes | single | no(paid) | http://www.urkund.com/ |
| Copycatch by CFL Software | 2002 | I/E | yes | single | no(paid) | www.copycatchgold.com |
| Wcopyfind by Louis A. Bloomfield | 2004 | I/E | yes | single | no(free) | http://www.plagiarism.phys.virginia/ |
| EVE2 by Canexus | 2001 | E | yes | single | yes(paid) | www.canexus.com |
| GPSP by Gllat consulting service | 1999 | I | yes | single | yes(paid) | http://www.plagiarism.com/ |
| MOSS by Alex Aiken | 1994 | E | No | multiple | no(free) | http://theory.stanford.edu/ aiken/moss/ |
| Jplag by Lutz Prechelt et al. | 1997 | E | Yes | multiple | yes(free) | https://jplag.ipd.kit.edu/ |
| Copyscape by Indigo Stream Technologies Ltd | 2011 | E | yes | url | yes(free) | http://www.copyscape.com/ |
| DOC Cop | 2006 | E | ? | single | no(free) | www.doccop.com/ |
| Ephorus by Ephorus B.V. | ? | E | yes | single | no(paid) | http://www.ephorus.com/ |
| iThenticate by iParadigms, LLC | 1996 | E | yes | single | no(paid) | http://www.ithenticate.com/ |
| PlagiarismDetect | 2008 | E | yes | single | no(free) | plagiarismdetect.org/ |
| Exactus Like by Ilya Sochenkov et al. | 2016 | E | yes | single | no(free) | http://like.exactus.ru/index.php/en |
| DupliChecker | 2006 | E | yes | single | no(free) | www.duplichecker.com/ |
| Plagiarisma | ? | E | yes | single | no(free) | http://plagiarisma.net/ |
| PlagiarismChecker by Darren Hom | 2006 | E | yes | single | no(free) | http://www.plagiarismchecker.com/ |
| Plagium by Septet Systems Inc. | 2006 | E | yes | single | no(free) | http://www.plagium.com/ |
| PlagTracker by Svetlana et al. | 2011 | E | yes | single | no(paid) | http://www.plagtracker.com/ |
| Quetext | ? | I/E | yes | single | no(free) | http://www.quetext.com/ |
| Turnitin by iParadigms | 2000 | E | yes | single | no(paid) | http://www.turnitin.com/ |
| Viper by All Answers Limited | 2007 | E | yes | single | yes(free) | http://www.scanmyessay.com/ |
| Maulik by Urvashi Garg et al. | 2016 | E | | single | no | Not available |
| Plagiarism Scanner | 2008 | E | yes | single | no(paid) | http://www.plagiarismscanner.com/ |
| Hawk Eye by Karuna Puri et al. | 2015 | E | yes | image | | Not available |
| Code Match by S.A.F.E | ? | E | yes | ? | yes(paid) | http://www.safe-corp.com/ |
| SID by Xin Chen et al. | 2004 | E | yes | single | yes(free) | http://software.bioinformatics.uwaterloo.ca/SID/. |
| SIM by D Gitchell | 1999 | E | No | multiple | yes(free) | http://www.cs.vu.nl/dick/sim.html |
| YAP3 by Michael J. Wise | 1996 | E | | multiple | no | Not available |
| PlagScan by PlagScan GmbH | 2015 | E | yes | single | no(free) | www.plagscan.com/ |

Note: ?: Need further research

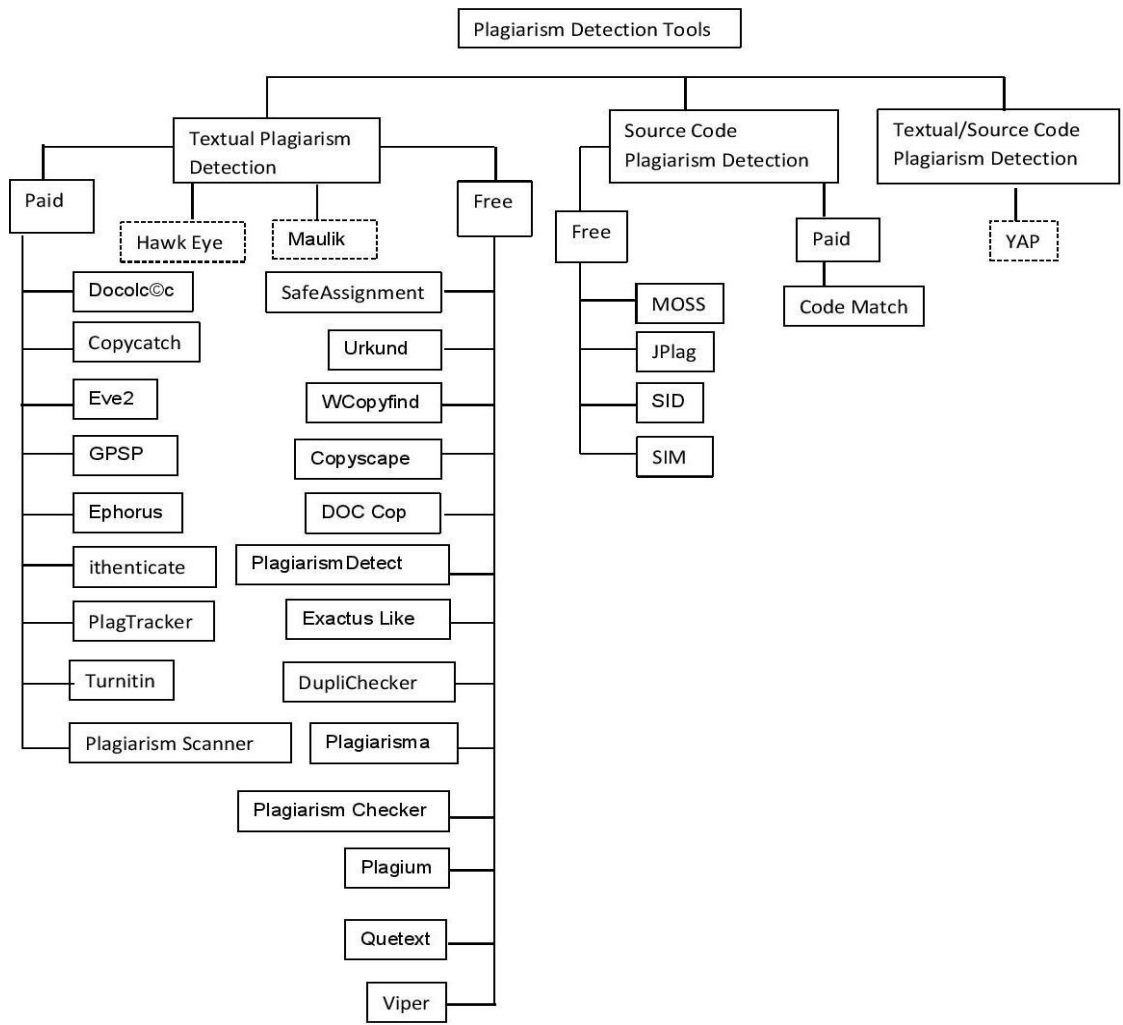

Figure 4: Classification of Plagiarism Detection Tools.

customer an opportunity to share plagiarism reports with other people by simply giving them the link, generated by this tool.

xxvi. **Hawk Eye [60]:** It is an innovative plagiarism detection system. This uses mobile scanner OCR(Optical Character Recognition) engine into convert image to text and that text it uses as input. The OCR Engine preprocess the clicked image in order to remove noise and disturbance from it and extract relevant keywords from image. The system uses plagiarism detection algorithms to remove unnecessary details like comments and changing variables names. or It uses string matching to detect plagiarism. It considers many limitations of existing well known plagiarism detection tools like Moss, JPlag, and Turnitin.

xxvii. **Code Match [8]:** Code Match compares source code and executable to detect plagiarism. It is developed by SAFE(Software Analysis and Forensic Engineering). It has also some additional functionality, which allows finding open source code within proprietary code, determining common authorship of two different programs, or discovering common, standard algorithms within different programs. It supports almost all existing programming languages.

xxviii. **SID-Software Integrity Diagnosis system [61]:** It detects plagiarism between programs by computing the shared information. It uses a metric in measuring the amount of shared information between two computer programs, to enable plagiarism detection and the metric is approximated by a heuristic compression algorithm. SID works in two phases. In the first phase, source programs are parsed to generate token sequences by a standard lexical analyser. In the second phase, Token Compress algorithm is used to compute heuristically the shared information metric d(x; y) between each program pair within the submitted corpus. Finally, all the program pairs are ranked by their similarity distances.

xxix. **SIM [62]:** This tool is to measure similarity between two C programs. It is useful for detection of plagiarism among a large set of homework programs. This tool is robust to common modifications such as name changes, reordering of statements and functions, and adding/removing comments and white spaces.

xxx. **YAP3[63]:** YAP is a system for detecting suspected plagiarism in computer program and other text submitted by the students. YAP3 is the third version of YAP which works in two phases. In the first phase, the source text is processed to generate token sequence. In second phase, each token is (non-redundantly) compared against all others strings.

xxxi. **PlagScan [64]:** PlagScan has separate packages for schools, universities and companies. To use this we need a paid account to open. It is not a free service but if someone does not like the service, membership cancellation is possible and money will be refunded.

## 5. Issues and Challenges

Based on our survey we observe that in past two decades, a large number of methods and tools have been developed to support fast and accurate plagiarism detection. Most prominent methods have been able to address the major issues related to (i) salient syntactic and semantic feature extraction, (ii) handling of both monolingual and cross-lingual plagiarism detection, and (iii) detecting plagiarism in both text data and program source code with or without using references. However, with the rapid growth of digital technology to support its reproduction, storage and dissemination, some important issues and research challenges are still left unattended. In this section, we highlight some of such issues and challenges that need to be addressed by computer science and linguistic researchers.

i. A detection method for both text data and source code that ensures both proof of correctness and proof of completeness is still missing, and hence an important issue.
ii. A proximity measure that guarantees detection of plagiarized text segment(s) in both intrinsic and extrinsic detection framework with high accuracy, is still not available.
iii. Developing a cross-lingual plagiarism checking tool that can perform without external references but ensures high accuracy is a challenging task.
iv. Developing a repository that maintains references based on author footprints, which is complete and accurate is another challenging task.
v. Developing a plagiarism checker that accepts an idea narrated by user and generates a detail plagiarism report (with similarity if detected from 1%-99%) with correct sources, is an important issue.

## 6. Conclusions

This paper has reported an exhaustive survey on plagiarism detection methods and tools in a systematic way. It has presented a taxonomy of various forms of plagiarism occur in text data and source code. Next, it has reported a large number of methods and tools under various categories and compared and analysed their pros and cons. Although in the past two decades, a large number of methods and tools have been introduced, we feel that there are still several issues and challenges left unattended. So, finally, we have highlight a list of issues and research challenges towards developing a plagiarism checker that is complete and correct for both monolingual and cross-lingual text data and for source code.